\documentclass[12pt]{article}
\usepackage[margin=1in]{geometry} 
\usepackage{graphicx} 
\usepackage{amsmath} 
\usepackage{amsxtra} 
\usepackage{amssymb} 
\usepackage{amsthm} 
\usepackage{latexsym} 
\usepackage{setspace} 
\usepackage{booktabs}
\usepackage{array}
\usepackage{threeparttable}
\usepackage[numbers]{natbib}

\theoremstyle{plain}
\newtheorem{assumption}{Assumption} 

\newtheorem{algorithm}{Algorithm} 

\begin{document}
	
	\title{Detecting Treatment Interference under the
		K-Nearest-Neighbors Interference Model} 
	
	\author{Samirah H.~Alzubaidi\footnote{Department of Mathematics, Al-Qunfudah University College, Umm Al-Qura University; E-mail: shzubaidi@uqu.edu.sa}, Michael J.~Higgins\footnote{Department of Statistics, Kansas State University; E-mail: mikehiggins@ksu.edu}}
	
	\maketitle

  \begin{abstract}
{  We propose a model of treatment interference where the response of a unit depends only on its treatment status and the statuses of units within its K-neighborhood. Current methods for detecting interference include carefully designed randomized experiments and conditional randomization tests on a set of focal units. We give guidance on how to choose focal units under this model of interference.  We then conduct a simulation study to evaluate the efficacy of existing methods for detecting network interference.
  We show that this choice of focal units leads to powerful tests of treatment interference which outperform current experimental methods.}
\end{abstract}

\section{Introduction}%
\label{makereference1}


Randomized experiments 
have long been viewed as the gold standard for causal inference~\citep{imbens2015causal}.
In epidemiology, researchers may want to study the effect of vaccines on a target population to protect individuals who are at risk of an infectious disease~\citep{hudgens2008toward}.
Technology companies such as Google, Amazon, Facebook, LinkedIn, Netflix, Twitter, and others run online randomized controlled experiments to evaluate the effect of a new feature or product on user engagement~\citep{gui2015network, pouget2019testing, saveski2017detecting}. 
However, in such settings, units under study may interact with each other; for example, a user assigned a new feature may interact with one not assigned the feature, thereby impacting the response of the latter user.
This interaction poses challenges in estimating and inferring treatment effects under traditional causal inference methodologies~\citep{sobel2006randomized}. 

In particular, a fundamental assumption in the traditional causal inference framework is that there is only a single version of each
treatment status and the response of a unit is unaffected by the treatment status of any other unit (see \citet{imbens2015causal} for a review).   
This is known as the \textit{stable unit treatment value assumption} (SUTVA)  \citep{rubin1980randomization}.
SUTVA is violated under settings in which there is \textit{treatment interference}---that is, when a treatment assigned to a unit affects the response of other units.
Effects on response due to treatment interference are also known as spillover, peer influence, social interaction, or network effects.

The dependence of a unit’s outcome on other units’ exposures or treatments poses statistical challenges because the \textit{potential outcome of a unit}---the hypothetical outcome of a unit given a realized treatment assignment---is not only affected by its own treatment status but also by the treatment conditions received by other units.
In some settings, interference can be considered as a nuisance parameter, and experiments may be designed in such a way to mitigate this interference, thereby reducing the bias in treatment effect estimates~\citep{eckles2016design}.
Although these designs may minimize the effect of interference, such designs are not always possible.
On the other hand, in other settings, estimating the causal effect in the presence of interference is of interest itself.
Examples of this include studies on the efficacy of vaccines in which vaccinated and non-vaccinated members of a population interact with each other and researchers are interested in the overall infection rates.   
Under these latter settings, considerable work has been devoted to the development of reasonable models of interference in order to ensure identification of both the direct effect of treatment and the effect of treatment spillover on the response~\citep{aronow2017estimating, forastiere2020identification, manski2013identification, sussman2017elements, toulis2013estimation}.  

In this paper, we introduce a model of treatment interference called the \textit{$K$-nearest neighbors interference model} (KNNIM).
Under KNNIM, the response of a unit is affected only by the treatment given to that unit and the treatment statuses of its $K$ nearest neighbors (KNN).
Such models of interference may be reasonable, for example, under social network settings, where only a few of the observable potential interactions (e.g.~accounts that a Twitter user follows) may be influential on a unit's response, and the strength of interaction may be measured by the amount of engagement between users.

We then perform a simulation study to determine how existing methods, and one newly developed method, for detecting treatment interference perform under data generated under a KNNIM model.
While these methods were originally developed to detect arbitrary interference  \citep{aronow2012general, athey2018exact, basse2019randomization, pouget2019testing, saveski2017detecting}, it is reasonable to assume that the efficacy of these methods may vary depending on the structure of interference.
However, little work has been done to assess how these methods perform under various interference models.
We repeatedly simulate data under a KNNIM model and apply these methods to the simulated data.
We then assess the power of these methods to successfully detect treatment interference when it is present and their likelihood of concluding insignificant interference when it is omitted.
Results suggest that methods which incorporate structured selection of focal units~\citep{aronow2012general, athey2018exact} tend to perform reasonably well on this type of data.
We then apply the existing methods to a study on the efficacy of an anti-conflict intervention in schools to determine their strength to detect interference on a real dataset.



The rest of this paper is organized as follows.
A motivating example is provided in Subsection \ref{makereference2.1}.
An overview on causal inference under interference is presented in Section \ref{makereference2}.
KNNIM is introduced in Section \ref{makereference3}.
Applying conditional randomization tests for detecting interference is discussed in Section \ref{makereference4}.
An algorithm on the selection of the focal units under KNNIM is provided in Section \ref{makereference5}.
Section \ref{makereference6} gives a summary of current methods of detecting interference.
Our proposed test statistic for detecting interference under KNNIM is given in Section \ref{makereference7}.
Section \ref{makereference8} evaluates current methods as well as our test under KNNIM model through a simulation.
The application of our method to our motivating example is given in Section~\ref{makereference11}.
Section~\ref{makereference10} concludes.

\subsection{Motivating Example: An Anti-Conflict Program in New Jersey Schools}
\label{makereference2.1}

To motivate our approach, we refer to a recent randomized field experiment 
assessing the efficacy of an anti-conflict intervention
aimed to reduce conflict among middle school students in 56 schools in New Jersey \citep{paluck2016changing}. 
In particular, the experiment was explicitly designed to determine whether benefits of the program can be propagated through social interactions between students.

The intervention was administered through ``seed'' students---those that are selected to actively participate and advocate for the anti-conflict program.
These students attended meetings with the program staff every two weeks to address conflict behaviors in their schools and to talk about strategies to mitigate peer conflict.
Additionally, seed students were encouraged to publicly reflect their opposition to conflict in their school---for example, identifying a common conflict in their school and creating a hashtag about it---and were also asked to distribute orange wristbands with the intervention logo to students that demonstrate anti-conflict attitudes.

Seed students were randomly assigned as follows.  
First, within each of the 56 schools, between 40 and 64 students were identified as being eligible to be seed students.
Then, from the 56 schools in the study, 28 schools were randomly assigned to receive the anti-conflict program.
Finally, within each of these assigned schools, half of the eligible students were selected to be seed students.
Analysis was performed only on students that were eligible to be seeds ($N =$ 2,451).

Of particular note, to assess potential pathways for treatment interference, students were asked to identify, in order, the 10 other students that they spent the most time with during the previous few weeks.
These students include both seed and non-seed students.
Specifically, the survey asks the following question: ``In the last few weeks I decided to spend time with these students at my school: (in school, out of school, or online) - Number 1 is for the person you spent most time with, then number 2, then number 3... You don't have to fill in all the lines! To make it easier, you can write down their initials here, then find their number. It can be boys and girls!''~\citep{paluck2016changing}.
Students' responses to this question may include both seed and non-seed students.
This yields a unique dataset in which the strength of the interaction between two individuals under study is explicitly recorded.
Hence, statistical analyses may benefit from an interference model, such as KNNIM, that allows for direct incorporation of the relative strengths of the interactions.
For this dataset, KNNIM models with $K$ up to 10 may be applicable.

An analysis performed by \citet{aronow2017estimating} estimated the indirect effect of being a seed student on wearing an orange wristband to be about 0.15  with a 95$\%$ confidence interval between about 8 and 23 percentage points.  
That is,
students exposed to treated peers were about 15\% more likely to report wearing an orange wristband in comparison to students in control schools.

\section{Background and Related Work}
\label{makereference2}
The Neyman-Rubin Causal Model (NRCM) is a popular model of response in causal inference~\citep{holland1986statistics, imbens2015causal, rubin1980randomization,  splawa1990application}.
Consider a simple experiment on $N$ units, numbered $1,\ldots, N$, where all units are given either a treatment or a control condition.
The NRCM assumes that the response of unit $i$, denoted $Y_i$ follows the model 
\begin{align*}
    Y_i = y_i(1)W_i + y_i(0)(1-W_i). 
\end{align*}
Here, $y_i(W_i)$ is the potential outcome under treatment status $W_i \in \{0,1\}$---the hypothetical response of unit $i$ had that unit received treatment status $W_i$---and $W_i$ is a treatment indicator: $W_i = 1$ if unit $i$ receives treatment and $W_i = 0$ if unit $i$ receives control.
Inherent in this model is the no interference assumption or \textit{stable unit treatment value assumption} (SUTVA).
This assumption states that there is only a single version of each treatment status and that a unit's outcome is only affected by its own treatment status and is not affected by the treatment status of any other unit~\citep{cox1958planning, rubin1980randomization}.

In many settings, SUTVA is not plausible, and considerable work has been performed on analyzing causal effects when SUTVA is violated. 
\citet{sobel2006randomized} showed that violating SUTVA can lead to wrong conclusions about the effectiveness of the treatment of interest.
\citet{forastiere2020identification} derive bias formulas for the treatment effect when SUTVA is wrongly assumed and show that the bias that is due to the presence of interference is proportional to the level of interference and the relationship between the individual and the neighborhood treatments.

When interference is present, the effect of a treatment on a unit's response may occur through direct application of the treatment to that unit, indirectly through application of treatment to units that interact with the original unit, or both~\citep{hudgens2008toward}.
We can extend the potential outcomes framework to account for both direct and indirect treatment components.  
Let $y_{i}(\mathbf W) = y_{i}(W_i, \mathbf W_{-i})$ denote the potential outcome of unit $i$ under treatment allocation $\mathbf W \in \{0,1\}^N$, where unit $i$ is given treatment $W_i$, and the remaining treatment statuses are allocated according to $\mathbf W_{-i}$.
Responses $Y_i$ satisfy
\begin{equation*}
    Y_i = \sum_{\mathbf w \in \{0,1\}^N} y_{i}(\mathbf w)\mathbf{1}(\mathbf W=\mathbf w),
\end{equation*}
where $\mathbf{1}(\mathbf W=\mathbf w)$ is an indicator variable that is equal to 1 if and only if the observed treatment status $\mathbf W$ is equal to the hypothetical treatment status $\mathbf w$.

The \textit{average direct effect} $\tau_{dir}$ is the average difference in a unit's potential outcomes when changing that unit's treatment status and holding all other units' treatment status fixed.
It may be defined as
\begin{equation}
\label{eq:deftaudir}
\tau_{dir} = \frac{1}{N} \sum_{i = 1}^N (y_i(1,\mathbf{1}) - y_i(0,\mathbf{1})),
\end{equation}
where $\mathbf{1}$ denotes a vector of all 1's.
In contrast to direct effect, the \textit{average indirect effect} $\tau_{ind}$ is defined as the average difference in a unit's potential outcome when changing all other treatment statuses from control to treated, holding its own treatment fixed.
It may be defined as
\begin{equation}
\label{eq:deftauind}
\tau_{ind} = \frac{1}{N} \sum_{i = 1}^N (y_i(0,\mathbf{1}) - y_i(0,\mathbf{0})),
\end{equation}
where $\mathbf{0}$ denotes a vector of all 0's.
The \textit{average total effect} $\tau_{tot}$ measures the average difference in potential outcomes between all units receiving treatment and all units receiving control:
\begin{equation*}
\tau_{tot} = \frac{1}{N} \sum_{i = 1}^N (y_i(1,\mathbf{1}) - y_i(0,\mathbf{0})). 
\end{equation*}
Summing~\eqref{eq:deftaudir} and~\eqref{eq:deftauind} yields the expression 
\begin{equation}
    \label{eq:sum}
    \tau_{tot} = \tau_{dir} + \tau_{ind}.
\end{equation} 

Alternatively, the quantities $\tau_{dir}$ and $\tau_{ind}$ may be defined respectively as\\ $\tau_{dir} = N^{-1}\sum_{i=1}^N (y_i(1, \mathbf{0}) - y_i(0, \mathbf{0}))$ and $\tau_{ind} = N^{-1}\sum_{i=1}^N (y_i(1, \mathbf{1}) - y_i(1, \mathbf{0}))$ while still ensuring that~\eqref{eq:sum} holds.  
These quantities may differ from~\eqref{eq:deftaudir} and~\eqref{eq:deftauind} if there is interaction between direct effects and indirect effects---that is, if the differences $y_i(1,\mathbf W_{-i}) - y_i(0,\mathbf W_{-i})$ differ depending on the allocation of treatment given to $\mathbf W_{-i}$.
Moreover, direct effects may be defined for each possible $\mathbf W_{-i}$---e.g.~$\tau_{dir}(\mathbf W_{-i}) = N^{-1}\sum_{i=1}^N (y_i(1, \mathbf W_{-i}) - y_i(0, \mathbf W_{-i}))$---however, such definitions may prevent a decomposition of the total effect into direct and indirect effects~\citep{hudgens2008toward}.
Finally, when SUTVA holds, $\tau_{tot} = \tau_{dir}$ and $\tau_{ind} = 0$.



There are a variety of strategies for designing and analyzing experiments  under treatment interference. 
One approach is to view interference as a nuisance parameter and to reduce the effect of treatment interference on causal estimates through effective experimental design.   
This line of work aims to use available information on potential interaction of units to design an experiment that mitigates the effect of this interaction.
Often, this is done through forming clusters with high within-cluster interaction and randomizing treatment across clusters rather than individual units~\citep{ eckles2016design, gui2015network, ugander2013graph}.
However, knowledge of the interaction network may not necessary to make progress on this problem---\citet{savje2021average} investigate methods for consistent estimation of treatment effects when the structure of interference is unknown. 
This approach may not be ideal when indirect effects are of interest to the researcher.

Rather than considering interference as a nuisance, some researchers tend to relax SUTVA and allow for different models of interference, considering interference effect as of primary interest.
One significant example of this involves experiments in the efficacy of vaccines where the likelihood of a person contracting an infectious disease depends on others in the same population who are vaccinated \citep{ halloran1995causal, hudgens2008toward, ross1916application}.
Under this setting, interference is allowed within groups but not across groups---this is referred to as a partial interference assumption \citep{sobel2006randomized}, i.e., SUTVA is assumed between groups~\citep{basse2018analyzing, hudgens2008toward, offer2021experimentation, rosenbaum2007interference, sobel2006randomized, tchetgen2012causal}.

A similar approach to partial interference assumes that treatment interference on a unit can only occur within a small closed neighborhood of that unit~\citep{sussman2017elements}---the $K$-nearest-neighbors interference model (KNNIM) introduced in this paper is a variant of this setting.  
Another common approach is to assume that the treatment condition can only ``spill over'' and affect the response of a control unit if a certain number or fraction of potential interactors of that unit receive treatment~\citep{gui2015network, toulis2013estimation}.
Finally, in its least restrictive form, \citet{aronow2017estimating} consider the use of Horvitz-Thompson estimators for estimating treatment effects under arbitrary forms of interference.

Another research direction focuses on the development of hypothesis tests to detect the presence of treatment interference in an experiment.
\citet{aronow2012general} introduces a framework for conditional randomization tests for detecting treatment interference.
\citet{athey2018exact} extend this approach to develop tests for more general forms of treatment interference.
\citet{basse2019randomization} build on this work and consider the validity of the test by conditioning on observed treatment assignment of the subset of units who received an exposure of interest.
\citet{saveski2017detecting} and \citet{pouget2019testing} develop an experimental framework to simultaneously estimate treatment effects and test whether treatment interference is present within an experiment.

\section{K-Nearest Neighbors Interference Model}
\label{makereference3}
To obtain meaningful estimates and inferences on treatment effects under interference, interference models often assume some kind of structure restricting how interference can propagate across units.
Otherwise, if a model allows for arbitrary interference, each unit will have a unique type of exposure depending on the treatment assignment for all ${N}$ individuals. 
This results in distinct $2^{N}$ potential outcomes for each unit and ${N}2^{N}$ potential outcomes for the experimental population in total.  
However, we only observe ${N}$ of these potential outcomes, and many causal quantities of interest will be unidentifiable under arbitrary interference.


Thus, the assumptions that researchers make about interference often lie strictly between assuming SUTVA and assuming arbitrary interference, and often greatly reduce the number of potential outcomes for each unit~\citep{aronow2017estimating, sussman2017elements, toulis2013estimation, ugander2013graph}. 
Many of these models specify that the units' outcomes are affected by the 
number/fraction of treated neighbors, but do not specify which neighbors impact unit response and how they affect the response. 


We now propose an interference model---the $K$-nearest-neighbors interference model (KNNIM)---where the treatment status of a unit $j$ can affect the response of a unit $i$ only if $j$ is one of $i$'s $K$--nearest neighbors.  
This model allows for neighbors of $i$ to contribute differing effects on the response of $i$ depending on the proximity of their relationship---neighbors that are ``closer'' to unit $i$ may have a larger influence on the response of $i$.
Additionally, this model restricts the number of potential outcomes to be $2^{K+1}$ for each unit. 


\subsection{Interaction Measure}
We begin formally introducing KNNIM by introducing an interaction measure $d(i,j)$ that measures how strongly unit $i$ associates with unit $j$.  
This measure does not necessarily need to be computed across every pair of units $(i, j)$; however, we assume that at least $K$ values of $d(i,j)$ can be computed for each unit $i$, $j\neq i$.   
Here, $d(i,j)$ may be measured explicitly.
For example Section~\ref{makereference2.1} describes an example where respondents assign numbers to 10 students, from 1 to 10, where 1 denotes the closest connection, 2 denotes the second closest connection, etc.~\citep{paluck2016changing}.
Alternatively, $d(i,j)$ may combine several interaction measures to form a proxy for overall interaction.
For example, an experiment on a social network may define $d(i,j)$ to be an index variable aggregating the number of comments, likes, and other forms of engagement performed by user $i$ and directed towards user $j$.
Smaller values of $d(i,j)$ may correspond to stronger or weaker interactions from $i$ towards $j$ depending on researcher preference.
In this paper, we assume smaller values correspond to stronger interactions.

Of particular note, the dissimilarity measure is allowed to be asymmetric; that is, $d(i,j)$ and $d(j,i)$ may differ.  
Such a property may be necessary if one user strongly influences another user, but not vice versa.
A common instance of this involves social media moguls; a mogul $i$ may induce strong engagement from millions of followers $j$, but may interact sparingly with the vast majority of these followers.  
This would suggest that followers of the mogul may be strongly impacted by an intervention given to the mogul---indicated by a small value of $d(j,i)$---but the mogul's behavior may not be altered by their followers---indicated by a large value of $d(i,j)$.

Additionally, it may also be the case that the same absolute value of $d(i,j)$ may be interpreted differently across users.
For example, suppose that $d(i,j)$ is an index variable for engagement on a social media platform.
If two users $i$ and $i'$ interact with the same user $j$ in identical ways, we may have $d(i,j) = d(i',j)$.
However, if $i$ engages with the platform often and $i'$ does so sparingly, then $d(i,j)$ may be relatively large for user $i$ (that is, $i$ may interact even more with close users $j^*$, leading to smaller values of $d(i,j^*)$),
but $d(i',j)$ may be relatively small for user $i'$.

\subsubsection{Remarks\label{sec:remarks}}


Note, when we define our interaction measure $d(i,j)$, we assume that these interactions can be measured precisely and without error.  
This assumption may be reasonable under certain settings---for example, the motivating example in Section~\ref{makereference2.1}---but may be unlikely to hold in others.
For example, although a social network may have an error-free record of interactions between users---and thus, it may be possible to exactly determine $d(i,j)$ on that network---an external observer of the network may only have a small fraction of these observations to determine the strength of interactions between users.
Moreover, even in the presence of perfect information, useful estimates and inferences still require careful selection of $d(i,j)$ to ensure it accurately measures the strength of the interaction between users. 
Settings under which these interactions are measured with error have been previously considered~\citep{savje2021average, leung2022causal}; such a consideration is outside of the scope of this paper but may be an area of further research.

Additionally, previous work on treatment interference has considered models where the interaction is determined by the absolute value of $d(i,j)$, rather than its value relative to $d(i,j^*)$ for other units $j^*$~\citep{leung2022causal}.
While such a model may be plausible under certain settings, the aforementioned examples suggest scenarios for which a model that relies on the relative value of $d(i,j)$ rather than its absolute value may be more appropriate.





\subsection{$K$-Neighborhood Interference Assumption}



Let $d(i, (j))$ denote the $j$th smallest value of $\{d(i,j^*), j^* \neq i\}$; that is, $d(i, (1)) < d(i, (2)) < \cdots$.
For ease of exposition, we assume that all values of $d(i,j)$ are unique (in practice, ties may be broken arbitrarily).
The \textit{$K$-neighborhood} of unit $i$, denoted $\mathbf{\mathcal{N}}_{iK}$, is the set of the $K$ ``closest'' units to unit $i$:
\begin{align*}
    \mathbf{\mathcal{N}}_{iK} = \{j : d(i,j ) \leq d(i, (K)), j = 1,2, \ldots ,K\}.
\end{align*}
Define $\mathbf{\mathcal{N}}_{-iK} = \{1,\ldots, N\} \setminus (i \cup \mathbf{\mathcal{N}}_{iK})$ as the set of units that are outside of $i$'s $K$-neighborhood.
Note that the sets
$\{i, \mathbf{\mathcal{N}}_{iK}, \mathbf{\mathcal{N}}_{-iK}\}$ form a partition of the $N$ units.


Recall that $W_i$ is a treatment indicator for unit $i$, and let $\mathbf{W} = (W_1, W_2, \ldots, W_N)$ = $\{W_{i}, \mathbf{W}_{\mathcal{N}_{iK}}, \mathbf{W}_{\mathcal{N}_{-iK}}\}$ denote the vector of treatment assignments given to all units $N$.
Additionally, recall that $y_{i}(\mathbf{W})$ denotes the potential outcome for unit $i$ under treatment allocation $\mathbf{W} \in \{0,1\}^N$.
Now we give the following assumption that defines the $K$-nearest neighbors interference model:


\begin{assumption}
\label{K-NIA2CH2}
($K$-Neighborhood Interference Assumption (K-NIA)). 
Units under study satisfy the $K$-Neighborhood Interference Assumption (K-NIA) if and only if, for each unit $i$ and for all treatment allocations $\mathbf{W}_{\mathcal{N}_{-iK}}$, $\mathbf{W}'_{\mathcal{N}_{-iK}}$, the potential outcomes satisfy,
\begin{align*} 
y_{i}(W_{i}, \mathbf{W}_{\mathcal{N}_{iK}}, \mathbf{W}_{\mathcal{N}_{-iK}}) = y_{i}(W_{i}, \mathbf{W}_{\mathcal{N}_{iK}}, \mathbf{W}'_{\mathcal{N}_{-iK}}).
\end{align*}
\end{assumption}

Assumption~\ref{K-NIA2CH2} states that the potential outcome of unit $i$ is only affected by its treatment and by the treatments assigned to its $K$-nearest neighbors.
Changing treatments for other units outside the $K$-neighborhood will not affect the potential outcome of unit $i$.
This is a special case of the neighborhood interference assumption (NIA) described in \citet{sussman2017elements}. 
In its most general form, the $K$-nearest neighbors interference model (KNNIM) assumes only that the treatment interference structure satisfies 
Assumption~\ref{K-NIA2CH2}.
For convenience, we will suppress the treatment statuses in  $\mathbf{W}_{\mathcal{N}_{-iK}}$ when referring to the potential outcomes $y_i$.


For ease of exposition, it is often convenient to view units under study as a mathematical graph.
For KNNIM, let $\mathbf{G}_{\text{KNN}} = (\mathbf{V}, \mathbf{E}_{\text{KNN}})$ denote a directed graph on $\|\mathbf{V}\| = {N}$ vertices; each vertex $i \in \mathbf{V}$ corresponds to a unit under study.
An edge $\vec{ij} \in  \mathbf{E}_{\text{KNN}}$ if and only if $j$ is one of $i$'s $K$--closest neighbors: that is, $j \in \mathcal N_{iK}$.
Note, by definition, $\vec{ii} \notin \mathbf{E}_{\text{KNN}}$.
Each edge $\vec{ij}\in  \mathbf{E}_{\text{KNN}}$ has weight equal to the interaction measure $d(i,j)$.
In this paper, we may refer to $\mathbf{G}_{\text{KNN}}$ as the \textit{weighted adjacency graph}.
Throughout this article, the terms vertex, unit, and individual will be used interchangeably.


Let $\mathbf{A}$ denote the $\mathbb{N}$ $\times$ $\mathbb{N}$ \textit{adjacency matrix} of $\mathbf{G}_{\text{KNN}}$, which indicates the presence or absence of an edge $\vec{ij}$ in the graph $\mathbf{G}_{\text{KNN}}$.  
That is, $A_{ij} = 1$ if $\vec{ij} \in \mathbf{E}_{\text{KNN}}$ and $A_{ij} = 0$ otherwise. 
Note that the diagonal elements of the adjacency matrix are zero; that is, $A_{ii} = 0$ for all $i$.

\subsection{Choosing the neighborhood size \textit{K}}
\label{makereference3.1.1}


The choice of $K$ for a given study may vary depending on the studies' field,
the purpose of the study, and the availability of data.
The experimenter may also use prior knowledge from previous studies to help choose $K$---for example, 
if previous studies have indicated that a person's behavior is influenced by their two closest friends, setting $K = 2$ may be appropriate. 
When possible, the $K$ should be selected in early phases of the study to help construct the adjacency matrix $\mathbf A$ when collecting data.

However, another factor that should be addressed when choosing the size of $K$ is the sample size needed to accurately quantify, estimate, and draw inference on the $K$-nearest neighbors indirect effects.
As mentioned above, number of possible exposures to treatments under KNNIM is $2^{K+1}$. 
Hence, to ensure sufficient power, many methods that incorporate KNNIM will require a sufficient number of units assigned to each of these exposure levels.
From our experience, a good heuristic is to require roughly 30 observations for each treatment exposure.
Under this heuristic, most studies may find models with $K = 2$ or $3$ to be most useful. 

Issues may arise if responses are used to inform the value of $K$.  
For example, a \textit{post-hoc} selection of $K$ could lead to inaccurate detection of treatment interference due to inherent multiple testing issues (inferences must account for testing both the appropriateness of $K$ and the presence of interference in the model) and/or bias in indirect effect estimates.   
It may be possible to incorporate additional structure into KNNIM to allow for a rigorous treatment of this problem, 
but such work is outside of the scope of this paper. 
See~\citet{alzubaidi2023estimation} for additional information about the estimation of indirect effects under KNNIM.


\section{Randomization Inference for Detecting Interference}
\label{makereference4}


We now describe the framework for randomization inference for testing the presence of treatment interference under KNNIM.
Recall that $\mathbf{W}$ is the treatment assignment vector and $y_{i}(\mathbf{W})$ is the potential outcome of unit $i$ under treatment $\mathbf{W}$. Let $T =T(\mathbf{W}, y(\mathbf{W}))$ denote a test statistic---a random variable where the randomness follows from the random treatment assignment vector $\mathbf{W}$. 
Let $\mathbf{W}^{obs}$ and $\mathbf{Y}^{obs}= \mathbf{Y}(\mathbf{W}^{obs})$ denote the observed treatment assignment vector and the observed outcome vector respectively.
Then, $T(\mathbf{W}^{obs}, \mathbf{Y}^{obs})$ is the observed value of the test statistic.
We aim to test the null hypothesis of no treatment interference for each unit 
\begin{equation}
\label{eq:nointhyp}
H_0: y_{i}(W_i,  \mathbf{W}_{\mathcal{N}_{iK}}) = y_{i}(W_i,  \mathbf{W}_{\mathcal{N}_{iK}}').
\end{equation}

Typically, randomization tests under the potential outcome framework assume a sharp null hypothesis of no unit-level treatment effects, and potential outcomes are able to be inferred under this sharp null across  randomizations~\citep{fisher1925statistical}.
However, since the hypothesis~\eqref{eq:nointhyp} does not make assumptions about direct effect of treatment on each unit, the potential outcome $y_{i}(W_i,  \mathbf{W}_{\mathcal{N}_{iK}})$ may not be imputable for randomizations under which $W_i \neq W_i^{obs}$.
Progress can be made by conditioning on a set of randomizations $\mathbf{\Omega}$ and choosing a test statistic $T$ such that $T$ is \textit{imputable} under randomizations in $\mathbf{\Omega}$~\citep{basse2019randomization}.  Afterward, a conditional $p$-value is obtained by computing, for example, the fraction of randomizations $\mathbf W' \in \mathbf \Omega$ such that
\begin{align*}
     |T(\mathbf{W}', y(\mathbf{W}'))| \geq |T(\mathbf{W}^{obs}, \mathbf{Y}^{obs})|.
\end{align*}

Following \citet{aronow2012general} and \citet{athey2018exact}, this conditional randomization inference can be performed by first selecting a subset of units under study called \textit{focal units} 
and then only considering randomizations of treatment $\mathbf W$  that do not affect the treatment status of the focal units. 
Only \textit{variant units}---those that are not focal units---can have differing treatment statuses across randomizations.
In other words,
we simulate draws from the random treatment assignment vectors conditional on the fixed treatment of the focal units.
Thus, the null hypothesis of no interference is sharp on the focal units since only treatment statuses of variant units---only those units that can impose indirect effects---are randomized.  
The test statistic $T$ is only computed on the outcomes of the focal units and hence, the test statistic is imputable under alternative treatment assignment vectors.

Randomization tests tend to be the preferred approach for testing for interference under the potential outcome framework.  
Asymptotic results for statistics for testing interference can be challenging to derive for a number of reasons, including having to account for inherent dependencies between units' treatment allocations induced through the adjacency matrix $\mathbf{A}$.
Hence, the use of asymptotic tests tends to be restricted either to settings that rely on strong distributional assumptions or for carefully designed studies.

Finally, while these approaches were originally developed for tests of treatment interference,~\citet{basse2019randomization} extend this work to build a framework for randomization tests for more general forms of causal effects. 

\section{Selection of the Focal Units}
\label{makereference5}

Although the choice of the focal units does not affect the validity of randomization tests for interference, it plays a key role in determining the power of these tests \citep{athey2018exact}.
More precisely, there is a trade-off between the size of the focal set (the set of focal units) and the size of the variant set (the set of variant units).
Adding additional focal units allows for larger sample sizes when testing for treatment interference---thereby increasing the power of these tests---but will decrease the number of potential randomizations on the variant units---which decreases their power. 
For general interference models, several useful heuristics for choosing focal units have been proposed, varying widely in complexity.  
We now outline a few of these methods.

The most basic approach, suggested by~\citet{athey2018exact}, is to simply select at random half of the units in the sample to be focal units---the other half are variant units.  
Note, this rule does not take into account, in any way, the interference model being assumed.

For models in which interference only exists between units with $d(i,j) \leq r$ (see Section~\ref{sec:remarks}),
\citet{aronow2012general} suggests a rule to ensure a significant amount of treated and control variant units within each focal units' neighborhood:
\begin{align*}
     N_{F} \in \arg \max_{N_{F}}(N_{F}E(N_{T, var, r})E(N_{C, var, r}))
\end{align*}
where $N_{F}$ is the number of focal units and $N_{T, var, r}$ and $N_{T, var, r}$ are the number of treated and control units in the variant set respectively within a ``distance'' of $r$ from a randomly selected focal unit. 

Finally, when the adjacency graph $\mathbf G = (\mathbf V, \mathbf E)$ is known,~\citet{athey2018exact} proposes using an $\varepsilon$--net as the set of focal units---a set of units such that there is path of $\varepsilon$ edges or fewer in $\mathbf G$ from any variant unit $j$ to some focal unit $i$~\citep{gupta2003bounded}. 
Note, this is equivalent to choosing a maximal independent set of units in the graph $G^\varepsilon = (\mathbf V, \mathbf E^\varepsilon)$---an edge $\vec{ij} \in \mathbf E^\varepsilon$ if and only if there is a path of $\varepsilon$ edges or fewer from $i$ to $j$ in $\mathbf G$.

Under KNNIM, we suggest choosing focal units in a way such that the $K$--neighborhoods of the focal units do not overlap.
This can be done by creating a $2$--net on the undirected adjacency graph $\mathbf{G^*_{\text{KNN}}} =  (\mathbf{V}, \mathbf{E}^*_{\text{KNN}})$--- an edge $ij \in \mathbf{E}^*_{\text{KNN}}$ if and only if $\vec{ij} \in \mathbf{E}_{\text{KNN}}$ and/or $\vec{ji} \in  \mathbf{E}_{\text{KNN}}$, where $\mathbf{E}_{\text{KNN}}$ is the edge set of the directed weighted adjacency graph $\mathbf{G_{\text{KNN}}}$.
The 2--net can then be used as the focal units.
This will enable us to remove dependencies between outcomes of focal units induced by indirect effects.
In fact, if treatment is Bernoulli-randomized across units, the responses of the focal units will be independent of each other.
Additionally, a substantial fraction of focal units may still be selected under this condition, increasing the power of the the randomization inference.



We now describe a simple algorithm to obtain a 2--net on the undirected adjacency graph $G^*_{\text{KNN}}$.


\begin{algorithm}\label{IndependentFocals}
Given a $K$-nearest neighbors undirected adjacency graph $\mathbf{G}^*_{\text{KNN}} = (\mathbf V, \mathbf{E}^*_{\text{KNN}})$, the following algorithm will obtain a 2--net on $\mathbf G^*$.
\begin{enumerate}
 \item \textbf{Step 1:} (Initialize)  Let $\mathbf{U}$ = $\mathbf{V}$.  Initialize the set of focal units $\mathbf F = \emptyset$.  Initialize the set of variant units $\mathbf I = \emptyset$. 
 \item \label{alg1:Step2} \textbf{Step 2:} (Select focal unit) While $\lvert \mathbf{U}\rvert > 0$, choose one vertex $i \in \mathbf U$ at random.  Set $i$ as a focal unit: $i \in \mathbf F$. 
 \item \textbf{Step 3:} (Find nearest neighbors) Set $\mathbf {I}$ equal to all units $j$ such that $ij \in \mathbf{E}^*_{\text{KNN}}$.
 \item \textbf{Step 4:} (Find neighbors of neighbors) Find all units $k \in \mathbf V \setminus \mathbf {I}$ such that, for some unit $j \in \mathbf {I}$, $jk \in (\mathbf E^*_{\text{KNN}})^{2}$.  Set these units $k \in \mathbf{I}$.
 \item \textbf{Step 5:} (Remove units) Remove all vertices in $\mathbf{F}$ and $\mathbf{I}$ from $\mathbf{U}$.
 \item \textbf{Step 6:} (Repeat or terminate) If  $\lvert \mathbf{U}\rvert = 0$, stop.  The set of focal units $\mathbf F$ is a 2--net for $\mathbf G^*_{\text{KNN}}$.  Otherwise, set $\mathbf{I} = \emptyset$ and return to Step~\ref{alg1:Step2}.
 \end{enumerate}
\end{algorithm}

\section{Current Methods for Detecting Interference}
\label{makereference6}

Current methods for detecting interference include conditional randomization tests~\citep{aronow2012general, athey2018exact} (as outlined in Section~\ref{makereference4}) and carefully designed experiments performed with the intention to detect interference~\citep{pouget2019testing, saveski2017detecting}.
We now provide a summary of these methods for testing for interference.
For randomization tests, we focus on the choice of test statistic used.
For experimental design methods, we describe both experimental setup and the test statistic.

\subsection{Test Statistics for Randomization Tests}
\label{makereference6.1}
\citet{aronow2012general} introduced the randomization inference approach for testing for interference between units, where units are affected by their own treatment and by the treatment assigned to their immediate neighbors. 
In this test, the treatment status for a subset of focal units remains fixed; the rest of the units are the variant subset.
The randomization inference is conditional on the observed treatment status of the fixed subset.  
That is, this test is on indirect effects resulting from the treatment allocation on the variant subset of units.
A variety of test statistics may be used under this framework.

The Pearson correlation coefficient $\rho$ between the outcomes of the fixed units ($\mathbf{Y_{F}}$) and the ``distance'' to the nearest unit of a particular treatment status in the variant subset  ($\mathbf{D}_{nearest}$) may be used as the test statistic:
\begin{align}
\label{eq:perasonrank}
\rho  = 
cor(\mathbf{Y_{F}},\mathbf{D_{nearest}}).
\end{align}
A common choice of distance is the Euclidean distance between pretreatment covariates.
This distance can be incorporated into the KNNIM framework through the interaction measure $d$.
\citet{aronow2012general} advocates for computing Pearson correlation coefficient on the ranks of these quantities; however, preliminary simulations suggest that the statistic $\rho$ tends to be more powerful for the models considered in Section~\ref{makereference8}. 


\citet{athey2018exact} extend this work and develop tests for more general realizations of interference (e.g. no higher-order interference).
As part of this work, they suggest additional test statistics for detecting interference.
The edge-level contrast statistic $T_{elc}$---a modification of a test statistic proposed by \citet{bond201261}---is the difference between the average outcomes of the focal units with treated neighbors and the focal units with control neighbors.  
Here, $T_{elc}$ averages over edges $ij$ where $i$ is a focal unit and $j$ is not a focal unit: 
\begin{align*} 
T_{elc} = \frac{\sum_{i,j\neq i}F_{i}A_{ij}(1-F_{j})W_{j}Y^{obs}_{i}}{\sum_{i,j\neq i}F_{i}A_{ij}(1-F_{j})W_{j}} - \frac{\sum_{i,j\neq i}F_{i}A_{ij}(1-F_{j})(1-W_{j})Y^{obs}_{i}}{\sum_{i,j\neq i}F_{i}A_{ij}(1-F_{j})(1-W_{j})},
\end{align*} 
where $F_i$ is an indicator variable satisfying $F_i = 1$ if and only if $i \in \mathbf F$.

A second test statistic is the score test statistic $T_{score}$~\citep{athey2018exact}.
This statistic is motivated by a model of treatment interference in which the indirect effect is proportional to the fraction of treated neighbors~\citep{manski1993identification,manski2013identification}.
The score test begins by computing 
\begin{equation*}
    r_i = Y_i^{obs} - \overline{Y}^{obs}_{F,0} - (\overline{Y}^{obs}_{F,1} - \overline{Y}^{obs}_{F,0})W_i,
\end{equation*}
for each focal unit $i \in \mathbf F$, where $\overline{Y}^{obs}_{F,1}$ and $\overline{Y}^{obs}_{F,0}$ are the average outcome for the treated and control focal units respectively.
Then, $T_{score}$ is the covariance between these $r_i$ terms and
\begin{equation}
    \frac{\sum_{j=1}^{N} A_{ij} W_{j}}{\sum_{j=1}^{N} A_{ij}},
\end{equation}
which is the fraction of treated neighbors for unit $i$.
This statistic is computed across only focal units that have at least one treated neighbor:
 \begin{align*} 
T_{score} = cov\left(r_i, \frac{\sum_{j=1}^{N} A_{ij} W_{j}}{\sum_{j=1}^{N} A_{ij}} \left |  F_{i} = 1, \sum_{j=1}^{N} {A_{ij}>0 }   \right. \right).
\end{align*} 




Finally, \citet{athey2018exact} consider the has-treated-neighbor test statistic $T_{htn}$, a modification of Pearson correlation coefficient~\eqref{eq:perasonrank}. 
Instead of using the distance to the nearest treated neighbor, this statistic uses an indicator variable $E_i$ for whether any of a unit's neighbors in the variant subset are treated: that is, $E_i = 1$ if and only if $\sum_{j}A_{ij}W_{j}(1-F_{j})>0$.
Then $T_{htn}$ is the correlation between this indicator and the outcomes for the focal units $\mathbf F$:
\begin{align*} 
T_{htn} = \frac{1}{S_{Y^{obs}_{F}}.S_{E}} \frac{1}{|\mathbf F|}\sum_{i\in \mathbf F}\left(Y^{obs}_{i} - \overline{Y}_F^{obs} \right)E_i, 
\end{align*} 
where $\overline{Y}_F^{obs}$ and $S_{Y^{obs}_{F}}$ are the sample mean and standard deviation of the outcomes for focal units respectively and $S_{E}$ is the sample standard deviation of the $E_i$ variables.

\subsection{Experimental Design Approach}
\label{makereference6.2}

\citet{saveski2017detecting} and \citet{pouget2019testing} present a two-stage experimental design to test for the presence of interference.  In this design, the units under study are divided into two groups and two experiments are performed simultaneously: for one group, treatment is assigned completely at random, and for another group, units are clustered and treatment is assigned across clusters rather than units.  
Then, estimates of the average direct effect are computed under the assumption of no interference for both the completely randomized and cluster randomized designs.
Finally, a standardized difference $T_{exp}$ is computed between these estimates:
\begin{align} 
\label{eq:definetexp}
T_{exp} = \frac{|\hat\tau_{cr} - \hat\tau_{cbr}|}{\hat\sigma_p},
\end{align} 
where $\hat\tau_{cr}$ and $\hat\tau_{cbr}$ are the estimates of the direct effect under the completely randomized and cluster randomized designs respectively and $\hat \sigma_p$ is a pooled standard deviation of responses from both the completely randomized and cluster randomized designs~\citep{saveski2017detecting}.
Large values of $T_{exp}$ imply the presence of indirect effects.  

A conservative test of the null hypothesis of no treatment interference can be performed at the $\alpha$ significance level by rejecting the null hypothesis if and only if $T_{exp} \geq \alpha^{-1/2}$.  
Additionally, as the number of units $n \to \infty$, it can be shown that $T_{exp}$ converges to a standard normal distribution (provided that cluster sizes remain fixed). Thus, an approximate size $\alpha$ test can be conducted by rejecting the null hypothesis of no interference if $T_{exp} \geq z_{1-\alpha/2}$, where $z_{1-\alpha/2}$ is the $1-\alpha/2$ quantile of the standard normal distribution.


\section{ K-Nearest Neighbors Indirect Effect Test Statistic}
\label{makereference7}

We now propose an additional test statistic designed to detect $K$-nearest neighbors indirect effects.
Let $\overline{Y}^{obs}(W_i,\mathbf{W}_{\ell= 1})$ and $\overline{Y}^{obs}(W_i,\mathbf{W}_{\ell= 0})$ denote the average response of observed units that are assigned to treatment status $W_i$ and have their $\ell$th nearest neighbor assigned to the treatment condition and the control condition respectively. 
The \textit{$K$-nearest neighbors indirect effect test statistic} $T_{knn}$ is obtained by computing differences in potential outcomes between focal units that receive the same treatment status but differ on the status of their $\ell$th nearest neighbor, and summing these differences across each of the $K$ nearest neighbors.

That is, for $W_i \in \{0,1\}$ and $\ell \in \{1, \ldots, K\}$, define
\begin{equation*} 
    T_{knn, \ell}(W_i) =  \overline{Y}^{obs}(W_i,\mathbf{W}_{\ell= 1}) - \overline{Y}^{obs}(W_i,\mathbf{W}_{\ell= 0}),
\end{equation*} 
and define $T_{knn, \ell}$ as a weighted average of these terms:
\begin{align*}
    T_{knn, \ell} = \frac{N_{Ft}}{|\mathbf F|}T_{knn,\ell}(1) +
\frac{N_{Fc}}{|\mathbf F|}T_{knn,\ell}(0),
\end{align*}
where $N_{Ft}$ and $N_{Fc}$  are the number of treated focal units and control focal units respectively.
 We then can define $T_{knn}$ as a sum of these $T_{knn,\ell}$ statistics:
\begin{align*} 
T_{knn} = \sum_{\ell=1}^K T_{knn,\ell}.
\end{align*} 


Note that, under the null hypothesis of no treatment interference, each of the $T_{knn,\ell}(W_i)$ terms should be close to 0.
Thus, since $T_{knn}$ is a linear combination of these terms, values of $T_{knn}$ that are relatively large in magnitude provide evidence against this null hypothesis, and so, $|T_{knn}|$ may be effective as a test statistic.
Additionally, note that the statistic $T_{knn,\ell}$ may be used  directly for a test of interference stemming from treatments assigned to the $\ell$th-nearest neighbor.

\section{Simulation}
\label{makereference8}

We now conduct a comparison and evaluate the performance of the methods covered in Section \ref{makereference6} and~\ref{makereference7} for testing the null hypothesis of no interference under the $K$-nearest neighbors interference model.

\subsection{Data Generation Procedure}
We generate the responses under the following model which satisfies KNNIM with $K = 3$:
 \begin{equation} 
 \label{definemodel}
Y_{i} = \mathbf{X}_{1} + \mathbf{X}_{2} + \mathbf{X}_{3} + \beta_{1}W_{i1} + \beta_{2}W_{i2} +\beta_{3}W_{i3} + \beta_{d}W_{i}.
\end{equation} 
In this model, we assume that the closest three neighbors affect the response $Y_i$;
we use $W_{i\ell}$ to denote the treatment status of the $\ell$th nearest neighbor of unit $i$.
The covariates $X_{j}$, $j = 1,2,3$, are independent and identically distributed $Normal(0,1)$ random variables.
We use the Euclidean distance between the covariates $\mathbf X_i$ and $\mathbf X_j$ as the interaction measure $d(i,j)$---units with more similar values of covariates are more likely to interact with each other.
Note that the model~\eqref{definemodel} defines the set of the potential outcomes for each unit $i$.
Simulated data is then generated by randomizing treatment across units.
Different models are obtained through varying the $\boldsymbol{\beta}=(\beta_{1}, \beta_{2}, \beta_{3}, \beta_{d})$ coefficients and the sample size $N$.
We consider sample sizes of $N=256$ and $N=1024$.


For each choice of sample size, we consider sixteen different models of interference.
We describe these models in Table~\ref{table1} in terms of the coefficients vector $\boldsymbol{\beta}$.
The first $3$ elements of $\boldsymbol{\beta}$ represent the indirect effect contributed by first, second, and third-nearest-neighbor respectively.
The last element $\beta_d$ is the unit's direct effect.
In all models considered, the closer the relationship to unit $i$, the greater the indirect effect: $|\beta_{1}|\geq  |\beta_{2}| \geq |\beta_{3}|$. 
The indirect effects in every set of three models represent the degree of interference starting from no interference in the first 3 models, followed by very weak interference in the second three models, weak interference in the next three models, moderate interference in the next three models, and finally strong interference in the last four models.     

For datasets with $N=256$ observations, 1,000 realizations of potential outcomes following each model are generated.
Tests of indirect effects are then applied to each of the 1,000 realizations.
Results for $N=256$ are given in Section~\ref{makereference9}.
Due to computational limitations, only 100 realizations are generated for models containing $N=1024$ units.
Results for $N=1024$ are given in the Supplementary Material.


\subsection{Simulation for Randomization Tests}
We compare the performance of both conditional randomization tests and experimental design approaches for detecting interference.
For the conditional randomization tests, for each set of generated potential outcomes, treatment is initially assigned completely at random to units, with half of the units receiving treatment and the other half receiving control.
Then, focal units are selected according to Algorithm~\ref{IndependentFocals}.
We then proceed with randomization tests as described in Sections \ref{makereference4} and \ref{makereference6.1}.
We evaluate the performance of the following test statistics: the Pearson correlation coefficient (Pearson)~\citep{aronow2012general}, the edge level contrast statistic (ELC), the score statistic (Score), the has-treated-neighbor statistic (HTN)~\citep{athey2018exact}, and the $K$-nearest neighbors indirect effect test statistic (KNN).

Test statistics are computed across 1,000 randomizations for each realization of the potential outcomes; 
for each randomization, treatment statuses are fixed for focal units and are completely randomized across variant units.
For each set of potential outcomes and for each choice of test statistic, we obtain a $p$-value for the null hypothesis of no treatment interference.
Thus, for $N=256$, we obtain a distribution of 1,000 $p$-values for each test statistic under each model.
The power of the tests can also be estimated by computing the fraction of $p$-values that fall beneath a pre-specified significance level $\alpha$.

\subsection{Simulation for Experimental Design Approach}

In addition, we follow the experimental design in~\citet{saveski2017detecting} (described in Section~\ref{makereference6.2}) to determine its efficacy for testing whether SUTVA holds under KNNIM.
For each set of generated potential outcomes, we divide the units into clusters of four units using a heuristic algorithm for the clique partitioning problem with minimum clique size requirement from~\citet{ji2004graph} (Algorithm 4).
This clustering is performed once per set of potential outcomes.

 We then randomly select half of the clusters to be cluster randomized; for this group, treatment is assigned at the cluster level, with half of the clusters receiving treatment and the other half receiving control.
For units belonging to the remaining clusters, each unit's cluster assignment is ignored, and treatment is completely randomized across all of these remaining units.
Again, half of these units receive treatment and the other half receive control.
For each set of potential outcomes, the random selection of clusters and the treatment randomization is performed 1,000 times. 

For each randomization, the statistic $T_{exp}$ in~\eqref{eq:definetexp} is computed.
We then perform a test of the null hypothesis of no treatment interaction at the $\alpha = 0.05$ significance level.  
A conservative test rejects this null hypothesis if $T_{exp} \geq \alpha^{-1/2}$ and an asymptotic test rejects the null if $T_{exp} \geq z_{1-\alpha/2}$.
Thus, for $N = 256$, we perform a total of 1,000,000 tests: that is, 1,000 tests for each of the 1,000 generated potential outcomes.
By computing the fraction of rejected null hypotheses, we are able to assess the Type I Error (Models 1--3) and the power (Models 4--16) of the experimental design approach.

\subsection{Discussion}
\label{makereference9}

\begin{figure}
\centering
\includegraphics[width=0.9\textwidth]{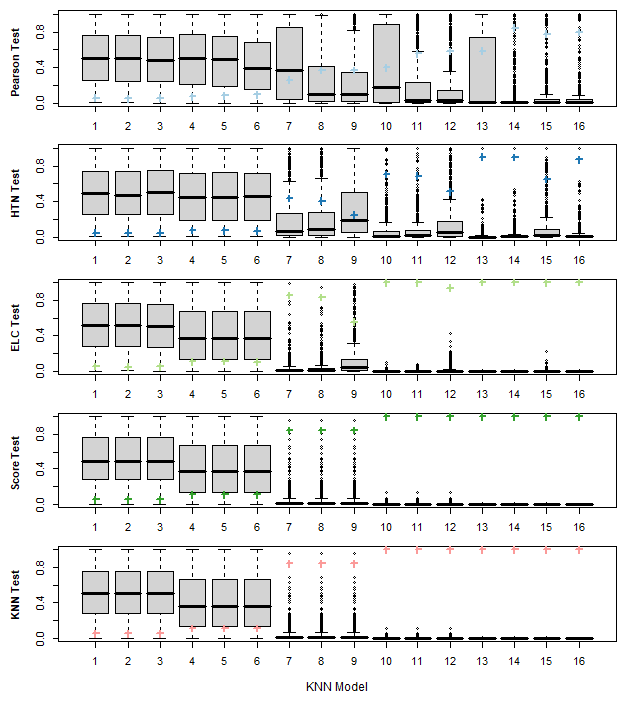}
\caption{Boxplots of $p$-values for the Pearson test (Pearson), has treated neighbor test (HTN), edge level contrast test (ELC),  score test (Score) and  $K$-nearest neighbors indirect effect test (KNN) under various KNNIM models.  
We use $N =256$ units and $K = 3$ nearest neighbors.  The $p$-values are estimated using 1,000 randomizations for each of the 1,000 generated potential outcome realizations.}
\label{figure2}
\end{figure}

Figure~\ref{figure2} provides a visual comparison of the distribution of $p$-values for the randomization tests to detect interference under KNNIM.
Table~\ref{table2.16} provides the estimated Type I Error and power of these tests (conducted at significance level $\alpha = 0.05$) across the 16 considered models.
As is expected by design~\citep{higgins2004introduction}, the $p$-values of all randomization tests under models without treatment interference (Models 1--3) are approximately distributed uniformly between 0 and 1.
All tests lack of power under very weak interference (Models 4--6) where the highest power is 0.110 for KNN test followed by 0.108 for Score test.
Under weak interference (Models 7--9), the ELC, Score, and KNN tests seem to outperform the Pearson and the HTN tests; the $p$-values are smaller overall for these three tests. 
Similar trends hold under moderate interference (Models 10--12) and strong interference (Models 13--16).
In particular, under strong interference, Score, KNN, and ELC tests have near 100\% power to detect treatment interference.  

However, the ELC and HTN tests seem to have some difficulty with detecting indirect effects when direct effects become large.
For example, the $p$-values for these three tests under Models 9 and 12---models that have comparatively larger direct effects---are substantially larger than under Models 7 and 8 and Models 10 and 11 respectively.  
The Score and KNN tests do not suffer from this loss of power as direct effects increase.  
For example, for Model 9, the Score and KNN tests have an estimated power 
of 0.844 and 0.839 respectively where the ELC and HTN tests have an estimated power of 0.553 and 0.249 respectively. 
Thus, for the considered tests, the Score and KNN tests seem to have the best combination of power in detecting treatment effects and isolating indirect effects in the presence of direct effects.
Similar comparisons between the methods hold for datasets with $N = 1024$ and/or when focal units are selected from only one treatment condition (see the Supplementary Material for details).

Figure~\ref{figure7} gives box plots of the estimated rejection rate across all 1,000 generated potential outcomes for both the conservative and asymptotic tests using the experimental design method~\citep{pouget2019testing, saveski2017detecting} with $N = 256$ and significance level $\alpha = 0.05$.
This plot also shows the estimated power of the considered randomization tests under these 16 models.
Table~\ref{table2.16} includes the median values of the rejection rates across the 1,000 generated potential outcomes for these tests.
The conservative experimental approach appears to lead to a very conservative test; the true Type I Error is much smaller than $\alpha = 0.05$, and the test appears to have weak power under very weak, weak and moderate interference. 
Even under Models 13--16, which exhibit strong interference, the conservative test only has a median power of approximately 0.6965.

The asymptotic test yields much more desirable results for our simulated data.
Overall, the Type I Error seems quite close to the nominal $\alpha = 0.05$.
The asymptotic test outperforms the Pearson and HTN randomization tests for almost all models of interference, and has a power close to 1 of detecting interference under Models 13--16.
However, the power of the asymptotic test still is behind that of the Score, KNN, ELC tests across all models.


When we increase the sample size to $N=1024$, the conservative approach seems to be powerful for moderate and strong interference while the asymptotic approach is powerful for all interference models except the very weak interference models.
However, both approaches remain comparatively less powerful than the Score, KNN, and ELC randomization tests (see the Supplementary Materials for details).

\section{Analysis of Anti-Conflict Program Experiment}
\label{makereference11}

In this section we reanalyze data from the motivating study described in Section~\ref{makereference2.1} designed to reduce conflict among middle school students in New Jersey.
Following \citet{paluck2016changing}, we only perform our analysis on seed-eligible students---hence, the adjacency matrix $\mathbf{A}$ only contains information about connections between seed-eligible students.
We then select a set of focal units following the procedure in Algorithm~\ref{IndependentFocals}.

For this study, randomization inference is then performed assuming complete randomization of treatment to the non-focal units.
Note, this is a simplification of how treatment was originally assigned to seed-eligible students---specifically, treatment was block-randomized with the schools serving as blocks.
However, as our focus is more on discussing the implementation of these randomization tests on data rather than confirming the results of~\citet{paluck2016changing}, we allow this simplifying assumption.

\subsection{Selecting \textit{K}}

Recall that the $K = 10$ closest connections were identified for each student.  
However, implementing a KNNIM model with $K = 10$ is impractical for this example.
For a study of this size ($N = $ 2,451), such a model would result in too many potential exposures for each unit (2,048 in total) to allow for meaningful inference to be performed on the indirect effect.
Moreover, seed-eligible students often identify connections with ineligible students which are not included in $\mathbf A$---in fact, most seed-eligible students have fewer than 3 connections with other seed-eligible students.  
This complicates the implementation of KNNIM with $K = 10$, which (from Section~\ref{makereference3}) is only well-identified when each observation $K$ has at least 10 connections.  

To determine whether a choice of $K$ is appropriate for this application, we first subset all seed-eligible students that have at least $K$ connections with other seed-eligible students.
We then calculate how many of these students are exposed to each of the $2^{K+1}$ treatment exposures.
Finally, we choose the largest $K$ that yields sufficient sample sizes (at least 30 students) for each exposure for our KNNIM model.

To make this explicit, suppose we consider a KNNIM model with $K = 2$.
This sample contains $N = 348$ units---that is, there are 348 seed-eligible students that interact with at least two other seed-eligible students.
Moreover, there are eight treatment exposures possible for each student in this sample; in Table~\ref{tableData1}, we see that each possible exposure has at least 34 students assigned to that exposure.
Hence, $K=2$ seems to be an acceptable choice.

Now, suppose we restrict our analysis further to only eligible students in treated schools who have at least $K$ = 3 seed-eligible nearest neighbors. 
In this case, the sample size is reduced to only 100 students.
Additionally, from Table~\ref{tableData2}, we see that there are an insufficient number of units assigned to each exposure---in fact, there is only one student in the sample that for which that student and all its three seed-eligible nearest neighbors are all treated. 
We conclude that $K=3$ yields an inappropriate model, and continue our analysis using a KNNIM model with $K = 2$.

\begin{table}[htb]
  \caption{Number of units in each exposure of Anti-Conflict Program Experiment with $K =2$ and $N = 348$}%
  \label{tableData1}
\centering
  \begin{tabular}[c]{lcccc}
  \hline
 \multicolumn{5}{c}{Indirect} \\
 \hline
        Direct & $(0,0)$&$(0,1)$&
        $(1,0)$&$(1,1)$ \\
\hline
       Treated & 38 & 42 & 39 & 34 \\
       Control & 40 & 59 & 46 & 50 \\
\hline
\end{tabular}
\end{table}

\begin{table}[htb]
  \caption{Number of units in each exposure of Anti-Conflict Program Experiment with $K =3$ and $N = 100$}%
  \label{tableData2}
   \centering
   \begin{tabular}[c]{lcccccccc} 
    \hline
    \multicolumn{9}{c}{Indirect} \\
   \hline
         Direct & $(000)$ & $(001)$&$(010)$&
         $(100)$&$(011)$&$(101)$&$(110)$&$(111)$ \\
 \hline
       Treated & 5 & 6 & 3 & 6 & 8 & 7 & 11 & 1 \\
       Control & 6 & 8 & 3 & 4 & 11 & 4 & 10 & 7\\
\hline
\end{tabular}
\end{table}



\subsection{Assessing indirect effects using randomization tests}
We evaluate the performance of the randomization tests for the following statistics: the Pearson statistic (Pearson), the edge level contrast statistic (ELC), the score statistic (Score), the has-treated-neighbor statistic (HTN), and the $K$-nearest neighbors indirect effect test statistic (KNN).
We choose focal units according to Algorithm~\ref{IndependentFocals} and treatment is re-randomized across non-focal units 1,000 times.
The $p$-value is the proportion of the replications where the absolute value of the simulated test statistic is greater than the absolute value of the observed test statistic.
Results are given in Table \ref{Data}.

\begin{table}[htb]
  \caption{Data Analysis of Anti-Conflict Program Experiment.}%
  \label{Data}
  \centering
  \begin{tabular}[c]{lc}
  \toprule
  
        Tests & $p$-value \\
\midrule
        Pearson & 0.72 \\
        ELC & 0.14 \\
        Score & 0.22 \\
        HTN &  0.45 \\
        KNN &  0.34 \\
        
          \bottomrule
\end{tabular}
\end{table}

For this modified experiment, all randomization tests fail to detect an indirect effect.  The $p$-value is smallest for the ELC test ($p = 0.14$), followed by the Score test ($p = 0.22$) and the KNN test ($p = 0.34$).

For context, an analysis of this experiment by~\citet{aronow2017estimating} estimated the indirect effect to be 0.154---that is, the probability that a non-seed student wears a wristband increases by about 15\% if they have a connection with a seed student.
Failure of these permutation tests to detect an indirect effect do not negate the findings of the original study.
For example, from Section~\ref{makereference8}, we find that permutation tests struggle to detect indirect effects of similar sizes consistently.
Additionally, this modified demonstration dramatically reduces the sample size of the original study, further decreasing the power of these tests.

\section{Conclusion}
\label{makereference10}
Traditional causal inference methodologies may fail to make reliable causal statements on treatment effects in the presence of interference. 
A substantial amount of recent work has been devoted to causal inference under interference, including methods for detecting treatment interference~\citep{aronow2012general, aronow2017estimating, athey2018exact, basse2019randomization, forastiere2020identification, manski2013identification, pouget2019testing, saveski2017detecting, sussman2017elements, toulis2013estimation}.

We consider a new model of treatment interference---the $K$-nearest-neighbors interference model (KNNIM)---in which the treatment status of a unit $i$ affects the response of a unit $j$ only if $i$ is one of $j$'s $K$ closest neighbors.  
We give advice for selecting focal units for conditional randomization tests for detecting interference under KNNIM, and suggest a new test-statistic---the \textit{$K$-nearest neighbors indirect effect test statistic} (KNN)---for these randomization tests.
We then perform a simulation study to compare the efficacy of both the randomization tests and experimental design approach for detecting interference under KNNIM.

Results suggest that randomization tests that incorporate our recommended selection of focal units tend to perform reasonably well on data satisfying KNNIM.  
Additionally, randomization tests using the score and KNN test statistics tended to be the most powerful for detecting interference, especially when direct effects are permitted to grow large relative to the indirect effects. 
Future research is needed to develop powerful tests under very weak interference.


\newpage

\newpage

\begin{figure}
\centering
\includegraphics[width=0.9\textwidth]{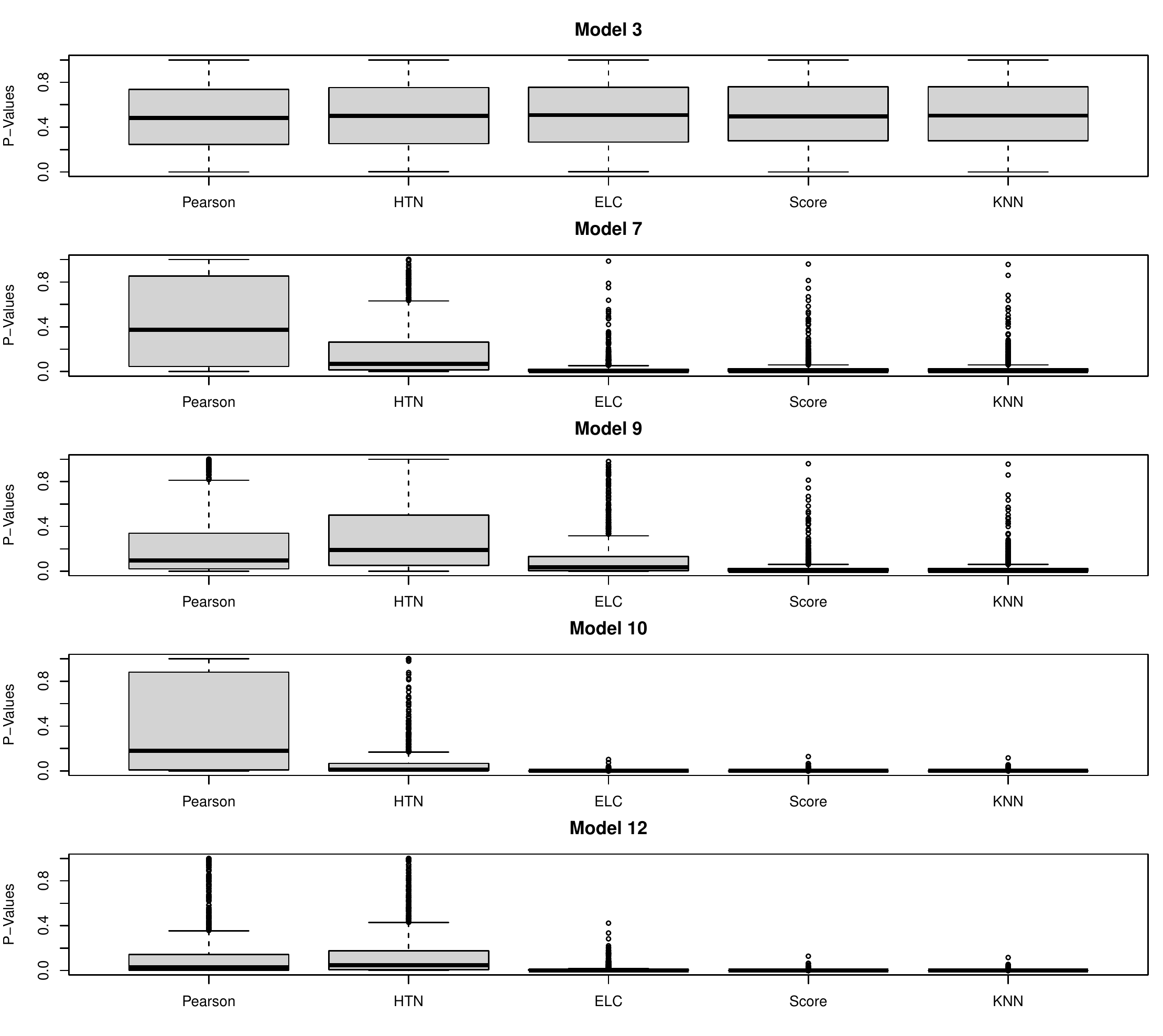}
\caption{Boxplots of $p$-values for the Pearson test (Pearson), has treated neighbor test (HTN), edge level contrast test (ELC),  score test (Score) and  $K$-nearest neighbors indirect effect test (KNN) under Models 3, 7, 9 10, and 12.  
We use $N =256$ units and $K = 3$ nearest neighbors.  The $p$-values are estimated using 1,000 randomizations for each of the 1,000 generated potential outcome realizations.}
\label{figurenew}
\end{figure}
\newpage

\begin{figure}
\centering
\includegraphics[ width=0.9\textwidth]{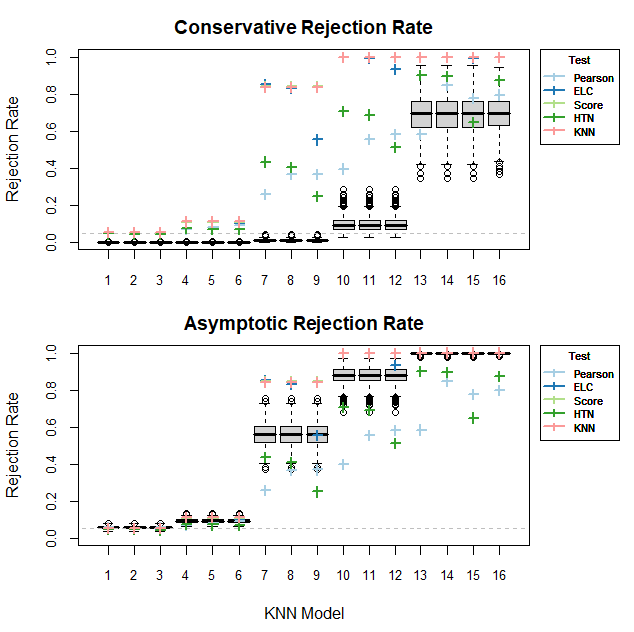}
\caption{ Boxplots of the estimated rejection rates under the experimental design approach for both the conservative and asymptotic tests of the null hypothesis of no treatment interference under various KNNIM models.  
Plots also contain the estimated Type I Error (Models 1--3) and power (Models 4--13) for the Pearson test (Pearson), edge level contrast test (ELC),  score test (Score), has treated neighbor test (HTN) and $K$-nearest neighbors indirect effect tests (KNN).
We use $N =256$ units and $K = 3$ nearest neighbors.  The rejection rates are estimated using 1,000 treatment assignments for each of the 1,000 generated potential outcomes.
Tests are performed at significance level $\alpha = 0.05$.}


\label{figure7}
\end{figure}
 
\newpage

\begin{table}[htb]
  \caption{Thirteen different interference models.}%
  \label{table1}
  \centering
  \begin{tabular}[c]{lc}\\
  \toprule
  
        Models & ($\beta_{1}$, $\beta_{2}$, $\beta_{3}$, $\beta_{d}$) \\
\midrule
        Model 1 & (0,0,0,0)\\
        Model 2 & (0,0,0,1) \\
        Model 3 & (0,0,0,4) \\
        Model 4 & (0.5,0.25,0.1,0)\\
        Model 5 & (0.5,0.25,0.1,0.3) \\
        Model 6 & (0.5,0.25,0.1,1) \\
        Model 7 & (2,1,0.5,0) \\
        Model 8 & (2,1,0.5,1) \\
        Model 9 & (2,1,0.5,4) \\
        Model 10 & (3,2,1,0) \\
        Model 11 & (3,2,1,1) \\
        Model 12 & (3,2,1,4) \\
        Model 13 & (30,20,10,0) \\
        Model 14 & (30,20,10,10) \\
        Model 15 & (30,20,10,40) \\
        Model 16 & (30,30,30,30) \\
  \bottomrule
\end{tabular}
\end{table}

\newpage

\begin{table}[htb]
  \caption{Estimated Type I Errors and power for tests of treatment interference for sample size $N=256$.}%
  \label{table2.16}
  \begin{center}
  \begin{tabular}[c]{lccccc|cc}\\
  \hline
  
        Models & Score & KNN & ELC & HTN & Pearson &   Cons & Asymp\\
\hline
        Model 1 & 0.050 & 0.051 & 0.053 & 0.045 & 0.055 & 0.000 &  0.056\\
        Model 2 & 0.050 & 0.051 & 0.046 & 0.044 & 0.049 & 0.000 &  0.056\\
        Model 3 & 0.050 & 0.051 & 0.048 & 0.043 & 0.055 & 0.000 &  0.056\\
        \hline
        Model 4 & 0.108 & 0.110 & 0.107 & 0.075 & 0.068 & 0.000 &  0.091\\
        Model 5 & 0.108 & 0.110 & 0.109 & 0.071 & 0.080 & 0.000 & 0.091\\
        Model 6 & 0.108 & 0.110 & 0.102 & 0.068 & 0.092 & 0.000 & 0.091\\
        Model 7 & 0.844 & 0.839 & 0.853 & 0.434 & 0.258 & 0.012 & 0.559\\
        Model 8  & 0.844 & 0.839 & 0.832 & 0.406 & 0.366 & 0.012 & 0.559\\
        Model 9  & 0.844 & 0.839 & 0.553 & 0.249 & 0.368 & 0.012 & 0.559\\
        Model 10  & 0.997 & 0.997 & 0.998 & 0.706 & 0.396 & 0.092 & 0.881\\
        Model 11  & 0.997 & 0.997 & 0.996 & 0.688 & 0.555 & 0.092 & 0.881\\
        Model 12  & 0.997 & 0.997 & 0.935 & 0.512 & 0.582 & 0.092 & 0.881\\
        Model 13  & 1.000 & 1.000 & 1.000 & 0.902 & 0.584 & 0.6965 & 0.998\\
        Model 14  & 1.000 & 1.000 & 1.000 & 0.897 & 0.846 & 0.6965 & 0.998\\
        Model 15  & 1.000 & 1.000 & 0.996 & 0.649 & 0.777 & 0.6965 & 0.998\\
        Model 16  & 1.000 & 1.000 & 1.000 & 0.874 & 0.796 & 0.6950 & 0.998\\
  \hline \\
\end{tabular}
\end{center}
    {\strut Estimated Type I Errors (Models 1--3) and estimated power (Models 4--16) for simulated data under KNNIM.
  Results are provided for the score test (Score), $K$-nearest neighbors indirect effect test (KNN), edge level contrast test (ELC), has treated neighbor test (HTN) and the Pearson test (Pearson).  Estimates of the median rejection rates under the experimental design approach for both the conservative (Cons) and asymptotic (Asymp) tests are also provided.
  We use $N =256$ units and $K = 3$ nearest neighbors. These values are estimated using 1,000 generated potential outcomes with 1,000 treatment assignments performed on each set of potential outcomes.
Tests are performed at significance level $\alpha = 0.05$.}
\end{table}

\newpage

\bibliographystyle{unsrtnat}
\bibliography{JCIbib}

\section{Supplementary Material }
\label{AppendixA}

\begin{figure}
\centering
\includegraphics[width=0.9\textwidth]{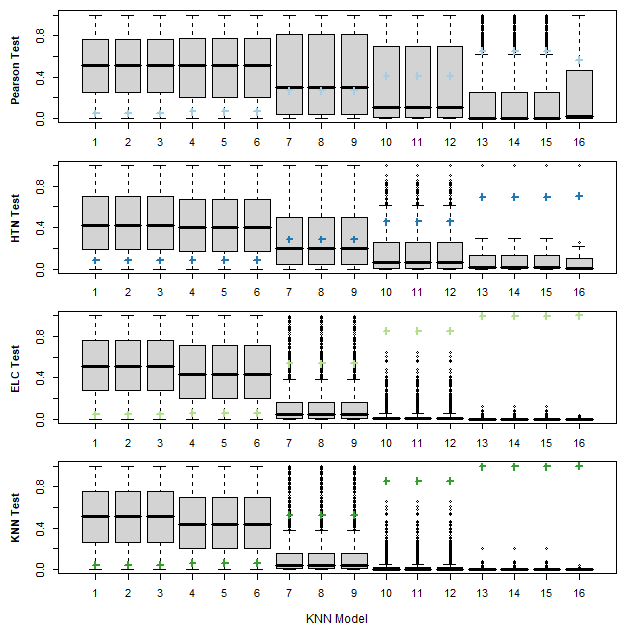}
\caption{Boxplots of $p$-values for the Pearson test (Pearson), has treated neighbor test (HTN), edge level contrast test (ELC) and  $K$-nearest neighbors indirect effect test (KNN) under various KNNIM models using only control focal units.  
We use $N =256$ units and $K = 3$ nearest neighbors.  The $p$-values are estimated using 1,000 randomizations for each of the 1,000 generated potential outcome realizations.
}
\label{figure1}
\end{figure}

\begin{figure}
\centering
\includegraphics[width=0.9\textwidth]{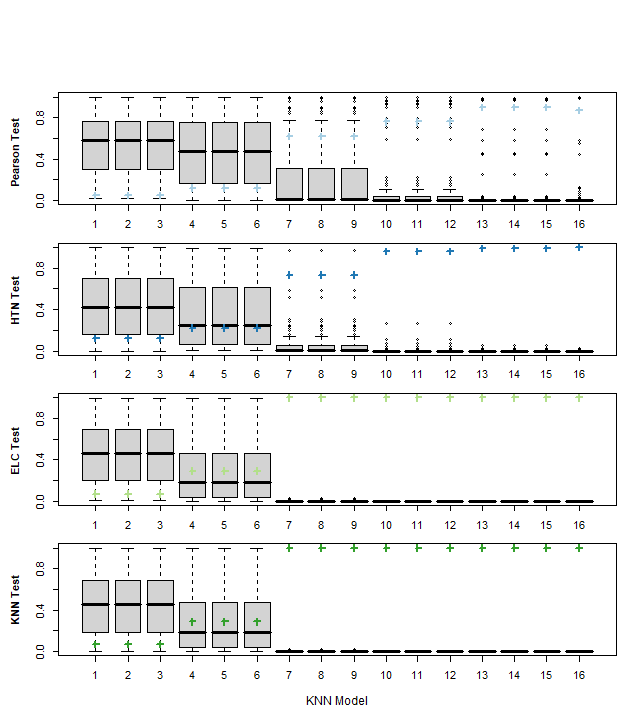}
\caption{Boxplots of $p$-values for the Pearson test (Pearson), has treated neighbor test (HTN), edge level contrast test (ELC) and  $K$-nearest neighbors indirect effect test (KNN) under various KNNIM models using only control focal units.  
We use $N =1024$ units and $K = 3$ nearest neighbors.  The $p$-values are estimated using 1,000 randomizations for each of the 100 generated potential outcome realizations.
.}
\label{figure3}
\end{figure}

\begin{figure}
\centering
\includegraphics[width=0.9\textwidth]{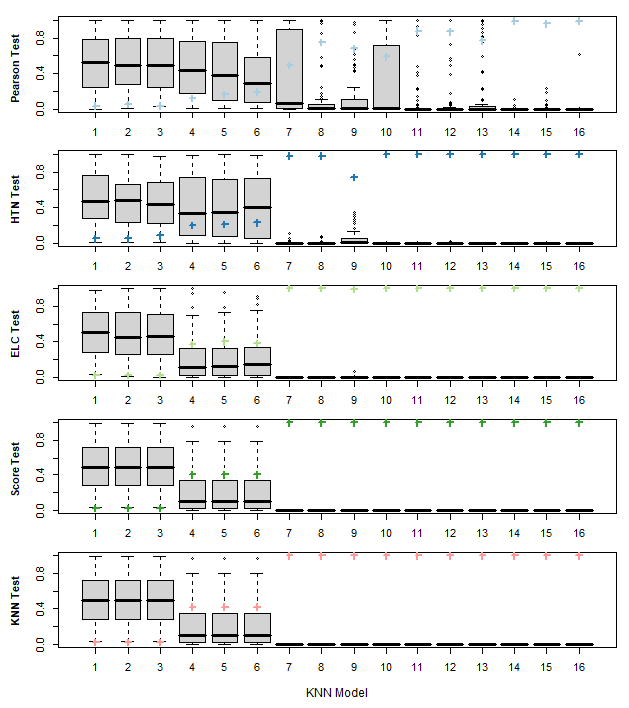}
\caption{Boxplots of $p$-values for the Pearson test (Pearson), has treated neighbor test (HTN), edge level contrast test (ELC),  score test (Score) and  $K$-nearest neighbors indirect effect test (KNN)  under various KNNIM models.  
We use $N =1024$ units and $K = 3$ nearest neighbors.  The $p$-values are estimated using 1,000 randomizations for each of the 100 generated potential outcome realizations.
}
\label{figure4}
\end{figure}



\begin{figure}
\centering
\includegraphics[width=0.9 \textwidth]{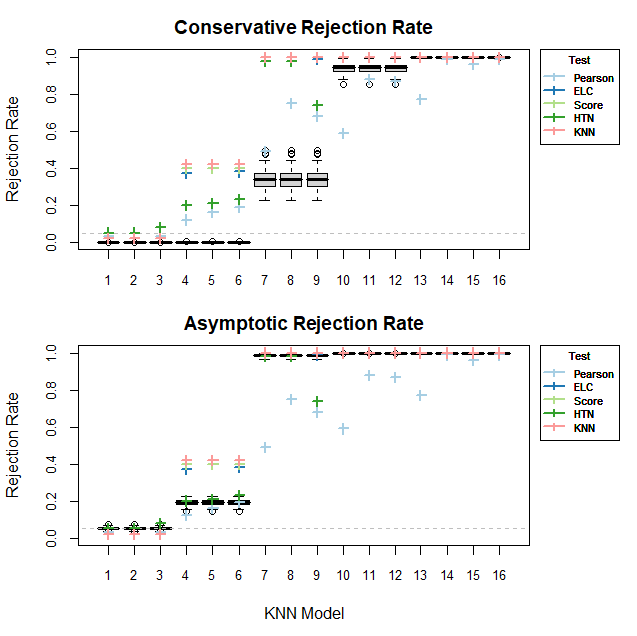}
\caption{ Boxplots of the estimated rejection rates under the experimental design approach for both the conservative and asymptotic tests of the null hypothesis of no treatment interference under various KNNIM models.  
Plots also contain the estimated Type I Error (Models 1--3) and power (Models 4--13) for the Pearson test (Pearson), edge level contrast test (ELC),  score test (Score), has treated neighbor test (HTN) and $K$-nearest neighbors indirect effect tests (KNN).
We use $N =1024$ units and $K = 3$ nearest neighbors.  The rejection rates are estimated using 1,000 treatment assignments for each of the 100 generated potential outcomes.
Tests are performed at significance level $\alpha = 0.05$.
}
\label{figure8}
\end{figure}

\begin{table}[htb]
  \caption{Estimated Type I Errors and power for tests of treatment interference for sample size $N=1024$.}%
  \label{table2.162}
  \begin{center}
  \begin{tabular}[c]{lccccc|cc}\\
  \hline
  
        Models & Score & KNN & ELC & HTN & Pearson &   Cons & Asymp\\
\hline
        Model 1 & 0.02 & 0.02 & 0.02 & 0.05 & 0.03 & 0.000 & 0.051\\
        Model 2 & 0.02 & 0.02 & 0.02 & 0.05 & 0.05 & 0.000 & 0.051\\
        Model 3 & 0.02 & 0.02 & 0.02 & 0.08 & 0.03 & 0.000 & 0.051\\
        \hline
        Model 4 & 0.40 & 0.42 & 0.37 & 0.20 & 0.12 & 0.000 & 0.192\\
        Model 5 & 0.40 & 0.42 & 0.40 & 0.21 & 0.16 & 0.000 & 0.192\\
        Model 6 & 0.40 & 0.42 & 0.38 & 0.23 & 0.19 & 0.000 & 0.192\\
        Model 7 & 1.00 & 1.00 & 1.00 & 0.98 & 0.49 & 0.339 & 0.986\\
        Model 8 & 1.00 & 1.00 & 1.00 & 0.98 & 0.75 & 0.339 & 0.986\\
        Model 9 & 1.00 & 1.00 & 0.99 & 0.74 & 0.68 & 0.339 & 0.986\\
        Model 10 & 1.00 & 1.00 & 1.00 & 1.00 & 0.59 & 0.943 & 1.00\\
        Model 11 & 1.00 & 1.00 & 1.00 & 1.00 & 0.88 & 0.943 & 1.00\\
        Model 12 & 1.00 & 1.00 & 1.00 & 1.00 & 0.87 & 0.943 & 1.00\\
        Model 13 & 1.00 & 1.00 & 1.00 & 1.00 & 0.77 & 1.00 & 1.00\\
        Model 14 & 1.00 & 1.00 & 1.00 & 1.00 & 0.99 & 1.00 & 1.00\\
        Model 15 & 1.00 & 1.00 & 1.00 & 1.00 & 0.96 & 1.00 & 1.00\\
        Model 16 & 1.00 & 1.00 & 1.00 & 1.00 & 0.99 & 1.00 & 1.00\\
  \hline \\
\end{tabular}
\end{center}
    {\strut Estimated Type I Errors (Models 1--3) and estimated power (Models 4--16) for simulated data under KNNIM.
  Results are provided for the score test (Score), $K$-nearest neighbors indirect effect test (KNN), edge level contrast test (ELC), has treated neighbor test (HTN) and the Pearson test (Pearson).  Estimates of the median rejection rates under the experimental design approach for both the conservative (Cons) and asymptotic (Asymp) tests are also provided.
  We use $N =1024$ units and $K = 3$ nearest neighbors. These values are estimated using 100 generated potential outcomes with 1,000 treatment assignments performed on each set of potential outcomes.
Tests are performed at significance level $\alpha = 0.05$.}
\end{table}

\end{document}